\documentclass[doublecol]{epl2}

\usepackage{amssymb}
\usepackage{amsmath}

\newcommand{\ket}[1]{\vert #1\rangle}

\newcommand{\an}[1]{\hat{#1}}
\newcommand{\cre}[1]{\hat{#1}^\dag}

\newcommand{\eh}{\textrm{e}}
\newcommand{\ii}{\mathrm{i}}
\newcommand{\diff}{\mathrm{d}}

\newcommand{\br}{\mathbf{r}}

\title{Phonon-induced artificial magnetic fields in optical lattices}

\shorttitle{Phonon-induced artificial magnetic fields}

\author{Alexander Klein\inst{1,2} \and Dieter Jaksch\inst{1,2}}
\shortauthor{A. Klein and D. Jaksch} %

\institute{
  \inst{1} Clarendon Laboratory, University of Oxford, Parks
Road, Oxford OX1 3PU, United Kingdom\\
  \inst{2} Keble College, University of Oxford, Parks Road, Oxford OX1 3PG, United Kingdom}


\abstract{We investigate the effect of a rotating Bose-Einstein
condensate on a system of immersed impurity atoms trapped by an
optical lattice. We analytically show that for a one-dimensional,
ring-shaped setup the coupling of the impurities to the Bogoliubov
phonons of the condensate leads to a non-trivial phase in the
impurity hopping. The presence of this phase can be tested by
observing a drift in the transport properties of the impurities.
These results are quantitatively confirmed by a numerically exact
simulation of a two-mode Bose-Hubbard model. We also give analytical
expressions for the occurring phase terms for a two-dimensional
setup. The phase realises an artificial magnetic field and can for
instance be used for the simulation of the quantum Hall effect using
atoms in an optical
lattice.}

\begin{document}

\maketitle

There are several phenomena in physics such as the quantum Hall
effect \cite{vonKlitzing-PRL-1980} or high-temperature
superconductivity \cite{Bednorz-ZPhysB-1986}, that have been known
for more than two decades, but where a full theoretical
understanding of the underlying mechanisms is still lacking. This is
due to the fact that a complicated many-body system consisting of
electrons in a crystal has to be described on a quantum mechanical
level, which quickly exceeds the capability of modern computers. As
Feynman suggested \cite{Feynman-IJTP-1982} it is therefore
worthwhile to search for alternative quantum systems that allow for
a clean realisation of the underlying models and that are easy to
manipulate. With such a quantum simulator, we would be able to do
simulations that by far exceed what is currently possible on
classical computers.

Ultracold atoms in optical lattices are a top candidate when it
comes to the simulation of condensed matter phenomena
\cite{Lewenstein-AiP-2006}. It has been shown theoretically that for
a suitable experimental setup the atoms in the lattice indeed
realise typical quantum Hall states
\cite{Sorensen-PRL-2005,Palmer-PRL-2006,Palmer-PRA-2008}. In a
solid-state setup, these states are realised by the application of a
strong magnetic field. Since the motional state of the
\emph{neutral} atoms in the optical lattice does not couple to a
magnetic field, the effect of this field has to be simulated by
other means. Proposals include using special lattice set\-ups and
Raman assisted hopping \cite{Jaksch-NJP-2003}, exploiting light with
orbital angular momentum
\cite{Juzeliunas-PRL-2004,Juzeliunas-PRA-2005}, or simply rotating
the lattice \cite{Cooper-PRL-2001}. Whereas the first proposals are
rather complicated to implement, the rotation of the lattice has
already been demonstrated in experiments \cite{Tung-PRL-2006}.
However, in order to observe the quantum Hall states the centrifugal
term caused by the rotation has to be carefully balanced by an
additional harmonic trapping potential confining the gas, which is
difficult to achieve in an experiment.

In this work, we therefore propose a way of exploiting a rotating
Bose-Einstein condensate (BEC) to create an artificial magnetic
field in an optical lattice system which is submerged into the
condensate. This has the advantage that the system in which the
quantum Hall states are to be observed, \textit{i.e.}~the optical
lattice, does not need to be rotated and thus there is no need of
employing a balancing potential. The BEC, on the other hand, only
provides the artificial magnetic field and we are not interested in
a direct observation of its states. Hence, we can employ a harmonic
trapping potential which overcompensates the centrifugal force
instead of exactly canceling it, which would not be possible if we
wanted to observe the quantum Hall states in the BEC directly.

The presence of the condensate can not only be used to provide an
artificial magnetic field, it also allows for the simulation of
other effects which are present in condensed matter physics
\cite{Bruderer-PRA-2007}. For example, it has been shown that for
suitably chosen parameters the coupling to the BEC phonons may
change the transport behaviour of the lattice atoms from coherent to
incoherent \cite{Bruderer-NJP-2008} and that an attractive
interaction between the lattice atoms can be mediated by the
condensate which leads to clustering effects \cite{Klein-NJP-2007}.
By including an artificial magnetic field, the whole setup can
therefore be used as a well-suited simulator for phenomena
encountered in condensed matter physics.

\section{Theoretical description}

In the following, we will derive an effective Hamiltonian for the
atoms in the optical lattice. A similar derivation has been given in
ref.~\cite{Bruderer-PRA-2007}, where it was assumed that the wave
function describing the condensate is real. In general, this will no
longer be the case if the condensate is rotated or translated,
leading to an additional phase in the hopping terms of the lattice
atoms. The Hamiltonian describing the whole system consists of three
parts, $\hat H = \hat H_B + \hat H_I + \hat H_a$. Here, $\hat H_a$
determines the free dynamics of the atoms of species $a$ trapped in
the optical lattice, which we will call impurities in the following.
In the laboratory frame, the BEC Hamiltonian $\hat{H}_B$ describing
the atoms of species $b$ and the density-density interaction
Hamiltonian $\hat{H}_I$ are
\begin{gather}
    \hat{H}_{B}=\int\!\diff\br\,\cre{\phi}(\br)\!\left[ \hat H_0
    +\frac{g}{2}\cre{\phi}(\br)\an{\phi}(\br)\right]\!\an{\phi}(\br)\,,\\
    \hat{H}_{I}=\kappa\int\!
    \diff\br\,\cre{\chi}(\br)\an{\chi}(\br)\cre{\phi}(\br)\an{\phi}(\br)\,,
\end{gather}
where $\an{\chi}(\br)$ is the impurity field operator,
$\an{\phi}(\br)$ is the condensate atom field operator satisfying
the usual bosonic commutation relations, and $\hat H_0$ contains the
kinetic energy term $-\hbar^2 \nabla^2/2 m_b$, an external trapping
potential $V_\mathrm{ext}(\br)$, the chemical potential $\mu$ and
any other terms that correspond to rotations or translations of the
BEC as will be detailed later. The coupling constants $g>0$ and
$\kappa$ account for the boson-boson and impurity-boson interaction,
respectively, and $m_b$ is the mass of a condensate atom.

In the tight-binding approximation, the impurity field operator can
be expanded as $\hat \chi(\br) = \sum_j \eta_j(\br) \hat a_j$, where
$\cre{a}_j$ creates an impurity atom in lattice site $j$ and
$\eta_j$ is the corresponding Wannier function
\cite{Jaksch-PRL-1998}. The probability densities of these functions
have a negligible overlap, \textit{i.e.}, $\int |\eta_j(\br)|^2
|\eta_{j'}(\br)|^2 \, \diff \br \approx 0$ for $j \neq j'$, which
will be important later. By using the above expansion for the field
operator $\hat \chi$ the Hamiltonian for the impurities is given by
$\hat H_{a} =
  -J_a \sum_{\langle i,j \rangle} \hat a^\dagger_{i} \hat a_{j} +
  (U_a/2) \sum_j \hat{n}_j(\hat{n}_j-1)
  - \mu_a \sum_{j}\hat n_j \,$ where $\mu$ describes the chemical
potential, $J_a$ is the hopping matrix element between two
neighbouring sites of the lattice, and $U_a$ is the on-site
interaction strength.

To investigate the influence of the rotating BEC on the impurities
we employ the Bogoliubov approximation to simplify the dynamics of
the condensate. To this end, we define $\hat \phi (\br) =
\phi_0(\br) + \delta\hat \phi(\br)$, where $\phi_0$ is a solution of
the Gross-Pitaevskii equation (GPE) \cite{Oehberg-PRA-1997} for
$\kappa = 0$,
\begin{equation}
   0 = (\hat H_0 + g |\phi_0(\br)|^2) \phi_0(\br) \,.
\end{equation}
In contrast to \cite{Bruderer-PRA-2007} we here do not assume that
the wave function $\phi_0$ is real. To proceed we require a weak
impurity-boson coupling $\kappa$ such that $|\kappa|/g
n_0(\br)\xi_h^D(\br)\ll1$, with $\xi_h(\br) = \hbar/\sqrt{m_b g
n_0(\br)}$ the healing length, $n_0(\br)=|\phi_0(\br)|^2$ the
density of the condensate and $D$ the number of spatial dimensions.
In this case, the deviation of $\an{\phi}(\br)$ from $\phi_0(\br)$
is of order $\kappa$,
\textit{i.e.},~$\langle\delta\an{\phi}(\br)\rangle\propto\kappa$,
where $\langle\,\cdot\,\rangle$ stands for the expectation value.
Inserting $\phi_0(\br) + \delta\an{\phi}(\br)$ into the Hamiltonian
$\hat{H}_{B} + \hat{H}_{I}$ and keeping terms up to second order in
$\kappa$, we obtain the linear term $\kappa\int
   \! \diff \br\,\cre{\chi}(\br)\an{\chi}(\br)[\phi_0(\br)\delta\cre{\phi}(\br) +
    \phi_0^\ast(\br)\delta\an{\phi}(\br)]$, in addition to the
standard constant and quadratic terms in $\delta\an{\phi}(\br)$ and
$\delta\cre{\phi}(\br)$.

The quadratic terms in $\delta\an{\phi}(\br)$ are diagonalised by
making use of the Bogoliubov transformation. The perturbation
$\delta\an{\phi}(\br)$ is expanded in terms of Bogoliubov phonons
$u_q$ and $v_q$, $\delta\an{\phi}(\br) = \sum_q^\prime[u_q(\br)\hat
b_q - v_q^\ast(\br) \hat b^\dagger_q]$, where $\hat b^\dagger_q$
creates a Bogoliubov phonon in mode $q$, the prime at the sum
indicates that the mode corresponding to the ground state is not
included, and the functions $u_q$ and $v_q$ solve the
Bogoliubov-de\,Gennes (BdG) equations
\begin{gather}
  \left[ \hat H_0 + g | \phi(\br)|^2
  \right] u_q(\br) - g \left(\phi(\br)\right)^2 v_q(\br) = E_q u_q(\br) \,,
  \\
  \left[ \hat H_0^\dagger + g | \phi(\br)|^2
  \right] v_q(\br) - g \left(\phi^\ast(\br)\right)^2 u_q(\br) = - E_q v_q(\br)
  \,.
\end{gather}
Putting this expansion for the condensate field operator into the
interaction Hamiltonian and using the fact that the overlaps of two
different impurity modes is negligible allows us to rewrite the
total Hamiltonian as a Hubbard-Holstein model
\cite{Mahan-2000,Holstein-Ann-1959}
\begin{equation}
\begin{split}
  \hat{H} =&\, \hat{H}_a +
    \sum_{j}\left.\sum_q\right.^\prime
      \hbar\omega_q(M_{j,q}\an{b}_q +
    M^\ast_{j,q}\cre{b}_q)\an{n}_j   \\
    &+\sum_j \bar{E}_j \,\hat{n}_j
    + \sum_q \hbar\omega_q \cre{b}_q\an{b}_q \,,
\end{split}
\end{equation}
with $\hbar\omega_q = E_q$ the energies of the Bogoliubov
excitations, the number operator $\hat{n}_j = \cre{a}_j\an{a}_j$,
the dimensionless matrix elements
$M_{j,q}=(\kappa/\hbar\omega_q)\int \!\diff\br\,
\left[\phi_0^\ast(\br)u_q(\br) - \phi_0(\br)
v_q(\br)\right]|\eta_j(\br)|^2$ and the mean field shift $\bar{E}_j
= \kappa \int\! \diff\br\, n_0(\br)|\eta_j(\br)|^2$.

The total Hamiltonian can be brought into a more intuitive form by
applying the unitary Lang-Firsov transformation \cite{Mahan-2000}
$\hat{H}_\mathrm{eff}=\hat{U}\hat{H}\hat{U}^\dagger$, with $\hat U =
\exp\big[
\sum_{j}\sum_q^\prime(M^\ast_{j,q}\cre{b}_q-M_{j,q}\an{b}_q)\hat{n}_j\big]$,
which yields
\begin{eqnarray}
\begin{split}
    \hat{H}_{\mathrm{eff}}= &\hat{U}\hat{H}_a \hat{U}^\dag +
    \sum_j (\bar{E}_j-E_j)\hat{n}_j - \sum_j
    E_j\hat{n}_j(\hat{n}_j -1) \\
    &-\frac{1}{2}\sum_{j\neq j'}V_{j,j'}\hat{n}_j\hat{n}_{j'} +
    \left.\sum_q\right.^\prime
    \hbar\omega_q\cre{b}_q\an{b}_q \,.
\end{split}
\end{eqnarray}
The presence of the BEC mediates a non-retarded interaction
$V_{j,j'} = \sum_q^\prime \hbar\omega_q \left(M_{j,q} M^\ast_{j',q}
+M^\ast_{j,q} M_{j',q} \right)$ between impurities in different
lattice sites $j$ and $ j'$. The characteristic potential energy of
an impurity in the deformed condensate is described by the polaronic
level shift $E_j = \sum_q^\prime \hbar\omega_q |M_{j,q}|^2$.

The transformation of the impurity Hamiltonian $\hat H_a$ yields
\begin{equation}\label{Eq:UHaU}
\begin{split}
  \hat{U}\hat{H}_a\hat{U}^\dagger =& -J_a  \sum_{\langle i,j
\rangle} (\hat X_i \hat a_i)^\dagger \hat X_j \hat a_{j} \\
&+ \frac{U_a}{2}  \sum_{j} \hat{n}_j (\hat{n}_j-1) -
    \mu \sum_{j} \hat{n}_j \,,
\end{split}
\end{equation}
where $\cre{X}_j =
\exp\big[\sum_q(M^\ast_{j,q}\cre{b}_q-M_{j,q}\an{b}_q)\big]$ is a
Glauber displacement operator that creates a coherent phonon cloud
around the impurity. Thus, the Hamiltonian $\hat H_\mathrm{eff}$
describes the behaviour of polarons according to an extended Hubbard
model \cite{Lewenstein-AiP-2006} provided that $c\gg a J_a/\hbar$,
with $c\sim\sqrt{g n_0/m_b}$ the phonon velocity and $a$ the lattice
spacing \cite{Bruderer-PRA-2007}.

In ref.~\cite{Bruderer-PRA-2007} it has been shown that for low
temperatures $k_B T \ll E_j$ the behaviour of the polarons is
essentially coherent. Here, $k_B$ is Boltzmann's constant. If also
$J_a \ll E_j$, we can treat the hopping term in eq.~(\ref{Eq:UHaU})
as a perturbation and trace over the condensate degrees of freedom
\cite{Bruderer-PRA-2007}. This allows us to derive the Hamiltonian
\begin{equation}
\begin{split}
  \hat H^{(1)}=&  \sum_{\langle i,j
\rangle} -\tilde J_{i,j} \,\cre{a}_i\an{a}_j +\frac{1}{2} \sum_j
\tilde U_j\hat n_j(\hat n_j-1)\\
 &
 - \sum_j \tilde{\mu}_j \hat n_j -\frac{1}{2}\sum_{i \neq j}V_{i,j}\,\hat n_i \hat n_{j}\,,
\end{split}
\end{equation}
with $\tilde{\mu}_j = \mu - \kappa n_0 + E_j$,
$\tilde{U}_j=U_a-2E_j$, and $\tilde J_{i,j} = J_a\,\langle\!\langle
\hat X_i^\dagger \hat X_j\rangle\!\rangle$. Here
$\langle\!\langle\,\cdot\,\rangle\!\rangle$ denotes the average over
the thermal phonon distribution. After some tedious algebra we find
\begin{align}
  \nonumber
  \langle\!\langle \hat X_i^\dagger \hat X_j\rangle\!\rangle =&\,
  \exp\left(-\frac{1}{2} \left.\sum_q\right.^\prime |M_{i,q} -
  M_{j,q}|^2 (2 N_q(T) +1) \right) \\
  &\times  \exp\left(
  \frac{1}{2} \left.\sum_q \right.^\prime M_{i,q}M_{j,q}^\ast -
  M_{i,q}^\ast M_{j,q} \right) \,,\label{Eq:HoppPhase}
\end{align}
where $N_q(T) =[\exp(\hbar \omega_q/k_B T)-1]^{-1}$ is the number
occupation of mode $q$. As was discussed in
ref.~\cite{Bruderer-PRA-2007}, the first exponential term leads to a
suppression of the coherent hopping $J_a$. The second exponential,
which does not occur for the cases presented in
\cite{Bruderer-PRA-2007} since there $\phi_0$ was assumed to be
real, has always an absolute value of 1 and causes a phase whenever
an atom hops from one lattice site $j$ to a neighbouring one $i$. By
defining the reduced hopping constant $\tilde J_a = J_a \exp\left(-
\sum_q^\prime |M_{i,q} -
  M_{j,q}|^2 (2 N_q(T) +1)/2 \right)$ the hopping term of the
Hamiltonian $\hat H^{(1)}$ can be written in the form $- \tilde J_a
\sum_{\langle i,j \rangle} \exp(2 \pi \ii \alpha_{i,j}) \hat
a^\dagger_i \hat a_j$ with $\alpha_{i,j} = 1/(4 \pi
\ii)\sum_q^\prime (M_{i,q}M_{j,q}^\ast - M_{i,q}^\ast M_{j,q})$.

This phase depends on the properties of the order parameter $\phi_0$
and drastically affects the dynamics of the impurities. For example,
for a quasi one-dimensional setup the presence of such a phase twist
in nearest neighbour hopping corresponds to a rotation of the ring
and thus causes persistent currents \cite{Nunnenkamp-PRA-2008}. This
is similar to the behaviour of an electron in a superconducting ring
subject to a magnetic flux. In a two-dimensional setup a suitable
phase factor realises an artificial magnetic field, which enables
for instance the investigation of quantum Hall effects
\cite{Palmer-PRL-2006}. In the remainder of this paper, we will
characterise the occurring phase term for different setups and
discuss ways of detecting its influence on the dynamics of the
impurity atoms.

\section{BEC in a ring}

\begin{figure}[t!]
\centering
  \includegraphics[width=4.3cm]{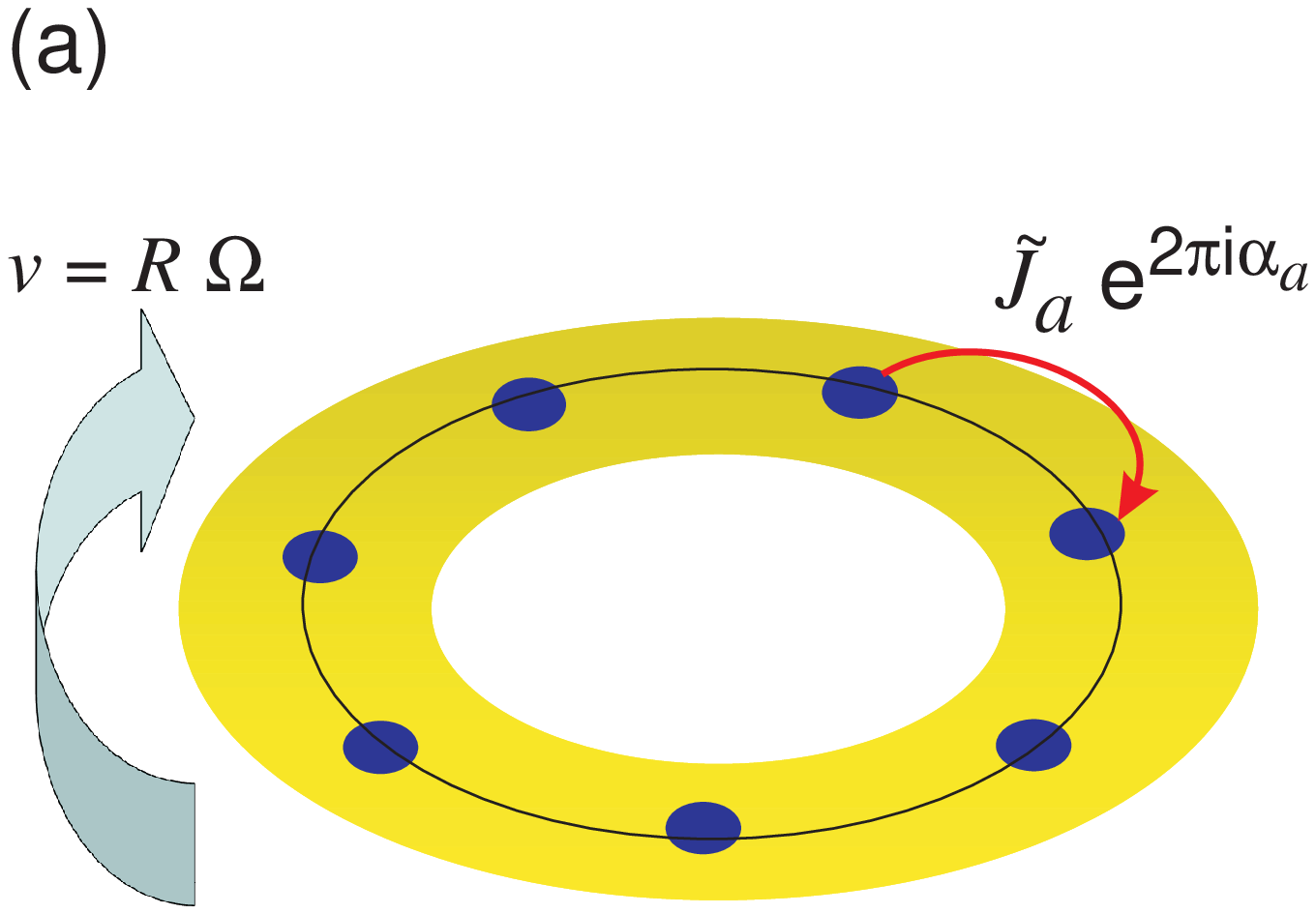}
  \includegraphics[width=4.3cm]{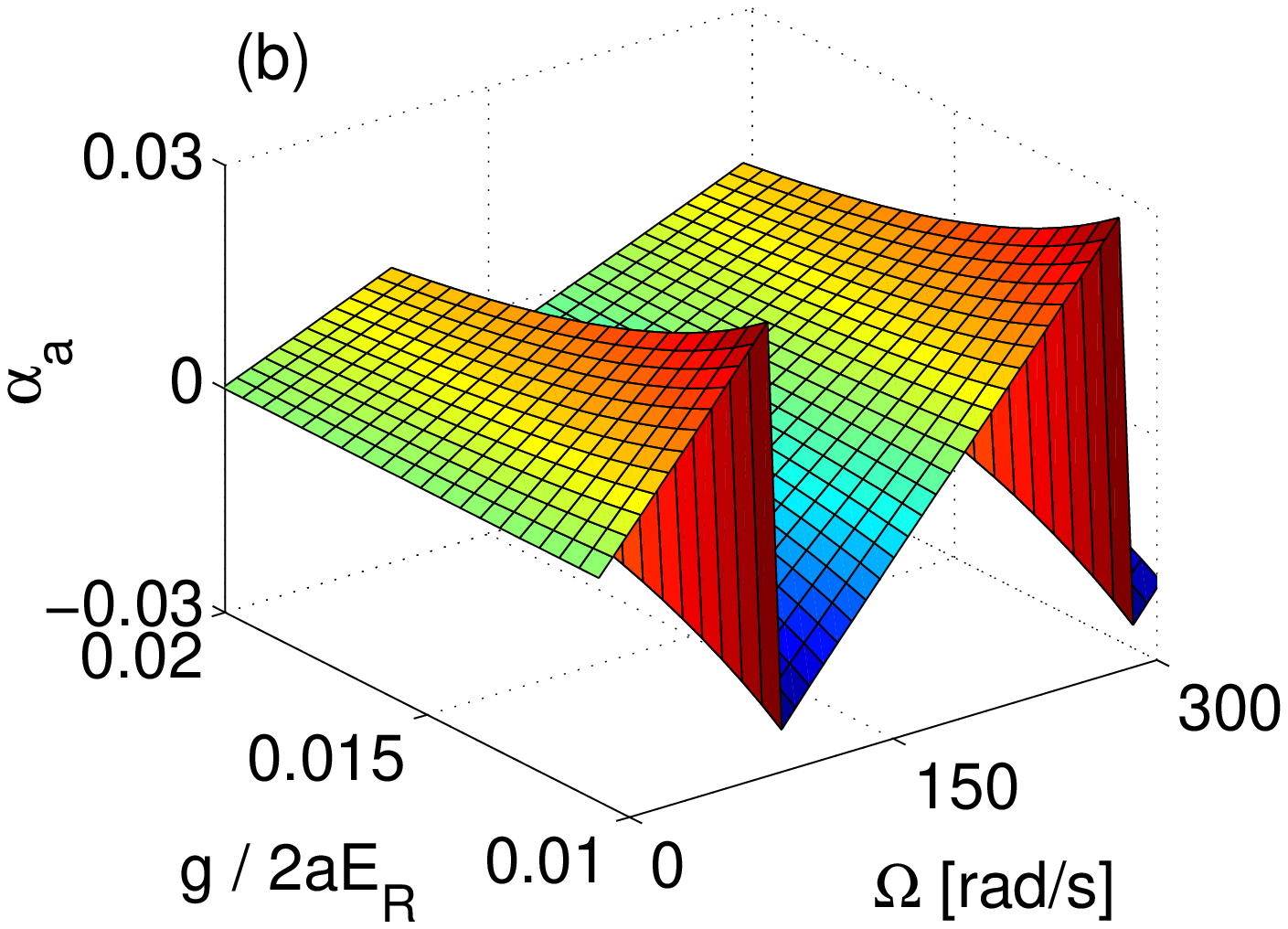}
\caption{\label{Fig:alpha1Dana} (a) Setup of the scheme. The ring
confining the BEC is rotated with angular velocity $\Omega$. This
induces a phase twist $\alpha_a$ when an impurity hops from one
lattice site (blue dots) to a neighbouring one. (b) Induced phase
twist $\alpha_a$ versus the angular rotation speed $\Omega$ and the
interaction within the BEC. We assumed a BEC of $^{87}$Rb with
linear density $n_0 = 5 \times 10^6\mathrm{m}^{-1}$ in a ring with a
circumference of $L= 12 \mu \mathrm{m}$. The impurities are
$^{21}$Na atoms in a lattice with 30 sites, which corresponds to a
lattice constant $a = 400\mathrm{nm}$. The coupling strength between
the impurity and the BEC is given by $\kappa/2aE_R = 0.035$, where
$E_R = (2 \pi \hbar)^2/2m_a (2a)^2$ is the recoil energy with $m_a$
the mass of the impurities. }
\end{figure}

We first consider the case where the BEC is loaded into a quasi
one-dimensional trap of length $L$ with periodic boundary
conditions, which corresponds to a ring of radius $R = L/2\pi$ and
has already been achieved in experiments
\cite{Gupta-PRL-2005,Heathcote-NJP-2008}. The impurities are
correspondingly trapped by a one-dimensional optical lattice which
is ring-shaped, see fig.~\ref{Fig:alpha1Dana}(a). We assume that the
ring trapping the condensate is rotated with an angular speed
$\Omega$. The Hamiltonian $\hat H_0$ describing the interaction-free
part of the BEC is then given by $\hat H_0 = -(\hbar^2/2m_b)
\partial^2/\partial x^2 + \ii \hbar v
\partial/\partial x -\mu$, where $v = R \Omega$. The full GPE is
solved by the wave function $\phi_0 = \sqrt{n_0} \exp(\ii q_0 x)$,
where $q_0 = 2 \pi j/L$ for some integer $j$ such that $q_0 = m_b
v/\hbar - \Delta q$ with $\Delta q \in [-\pi/L, \pi/L]$. This choice
of $q_0$ ensures that the BEC is in its ground state with a chemical
potential $\mu = \hbar^2 q_0^2/2 m_b + g n_0 - \hbar v q_0$ and that
the phase of the BEC fulfills the periodic boundary conditions.

The BdG equations are solved by the mode functions $u_q(x) = A_q
\exp[\ii (q+q_0)x]/\sqrt{L}$ and $v_q(x) = B_q \exp[\ii
(q-q_0)x]/\sqrt{L}$, where the coefficients are given by the
relation $A_q \pm B_q = (E^B_q/\varepsilon^0_q)^{\pm 1/2}$, and the
quasi-momenta fulfil the condition $q = 2 \pi j/L$ for some integer
$j$. The energies of the Bogoliubov modes are given by $\hbar
\omega_q = E^B_q - \hbar^2 q \Delta q / m_b$ with $E^B_q =
\sqrt{\varepsilon^0_q(\varepsilon^0_q+2gn_0)}$ and $\varepsilon^0_q
= \hbar^2 q^2/2m_b$. In order to calculate the coupling matrix
elements $M_{j,q}$ we assume that the lattice trapping the
impurities is sufficiently deep such that the Wannier functions
$\eta_j$ are well approximated by Gaussians of width $\sigma$
centered at lattice site $j$, \textit{i.e.},~$\eta_j(x) \approx
\exp[-(x-x_j)^2/2\sigma^2]/\sqrt{\sqrt{\pi}\sigma}$, where $x_j$ is
the position of lattice site $j$. We furthermore note that due to
the translational invariance the phase $\alpha_{i,i+1}=\alpha_a$
does not depend on the index $i$ and we thus find
\begin{equation}\label{Eq:alphaa_1D}
    \alpha_a = \frac{1}{2\pi}\frac{\kappa^2 n_0}{L}  \sum_{q \neq
    q_0}
  \frac{\varepsilon_q^0}{E_q^B} \frac{1}{(\hbar
  \omega_q)^2}\, \eh^{-q^2 \sigma^2/2} \sin\left( q a \right) \,.
\end{equation}
We see that the finite width of the Wannier functions $\sigma$ leads
to an effective cut-off of the sum for large quasi-momenta $q$ and
thus large energies $\hbar \omega_q$. Furthermore, the expression
for $\alpha_a$ contains the Bogolibov excitation energies $\hbar
\omega_q$ of the rotating condensate as well as the static structure
factor $\varepsilon_q^0/E^B_q$ of the \emph{non}-rotating BEC
\cite{Pitaevskii-2003,Griessner-NJP-2007}.

The value of $\alpha_a$ depends critically on the rotation speed
$\Omega$. For zero rotation it can be shown from
eq.~(\ref{Eq:alphaa_1D}) that $\alpha_a = 0$. If the rotation speed
is increased, the phase twist $\alpha_a$ initially increases
linearly with $\Omega$, as shown in fig.~\ref{Fig:alpha1Dana}(b).
However, for a critical rotation speed a jump occurs and the value
of the phase twist changes from positive to negative. This behaviour
can be explained as follows: Due to the finite size of the ring, the
possible quasi-momentum states of the BEC are restricted to values
$q_0 = 2 \pi j/L$ for some integer $j$. For rotation speeds close to
zero, the ground state of the BEC exhibits zero quasi-momentum. Only
if the rotation speed is above a critical value
$\Omega_\mathrm{crit} = \hbar /2 m_b R^2$ will the ground state
change to a non-zero quasi-momentum. Further quasi-momentum jumps
occur at odd multiples of this critical angular speed. Together with
the change in the quasi-momentum of the BEC, also the jump in the
induced phase twist $\alpha_a$ occurs. To some extend one could thus
say that the induced phase is caused by a mismatch of the angular
rotation speed $\Omega$ and the quasi-momentum $q_0$ of the BEC,
which is enforced to be quantised by the boundary conditions. It
should be noted that at the critical points where the BEC changes
its quasi-momentum, the ground state of the condensate is
degenerate, which leads to a failure of the Bogoliubov approximation
\cite{Rey-PRA-2007}.

The parameters presented in fig.\ref{Fig:alpha1Dana}(b) fulfill the
conditions for our derivation to be valid and can be achieved using
present experimental techniques
\cite{Bruderer-PRA-2007,Klein-NJP-2007,
Bruderer-NJP-2008,Madison-PRL-2001,Hodby-PRL-2001}. The precise
value of the coupling constant $g$ can be changed via a Fesh\-bach
resonance. The temperature chosen for our calculations is zero, but
the discussed effects will be observable as long as the temperature
is lower than the polaron energy, which is typically on the order of
a few tens of nanokelvin \cite{Bruderer-PRA-2007}. This also ensures
that the first factor in eq.~(\ref{Eq:HoppPhase}) is close to one
leading to sufficiently large hopping. In the above
examples the size of the induced phase twist can reach values as
large as $\alpha_a \approx 0.03$. This will significantly affect the
phase of the atomic wave function and make effects of the induced
artificial magnetic field observable even if the lattice only
consists of a few tens of sites in each direction.

\subsection{Drift of the impurities}

The presence of the phase twist $\alpha_a$ can be measured by
observing its influence on the transport properties of the
impurities. We consider the following case: A single impurity is
initially trapped in a finite spatial region of the lattice. This
can be achieved by applying an additional trapping potential
$V_\mathrm{ini}(j) = V_0 (j-j_0)^2$, in which case the single
impurity starts in the state $\ket{\psi}_\mathrm{ini} \propto \sum_j
\exp\left[-(j-j_0)^2/2w^2_\mathrm{ini}\right]\, \hat a^\dagger_j
\ket{\mathrm{vac}}$, where $w_\mathrm{ini}$ is the width of the
initial state. Without a phase twist $\alpha_a$ after turning the
initial potential off, the impurity expands symmetrically around the
lattice site $j_0$ and no overall drift to either higher or lower
lattice sites occurs.

\begin{figure}
\centering
  \includegraphics[width=4.3cm]{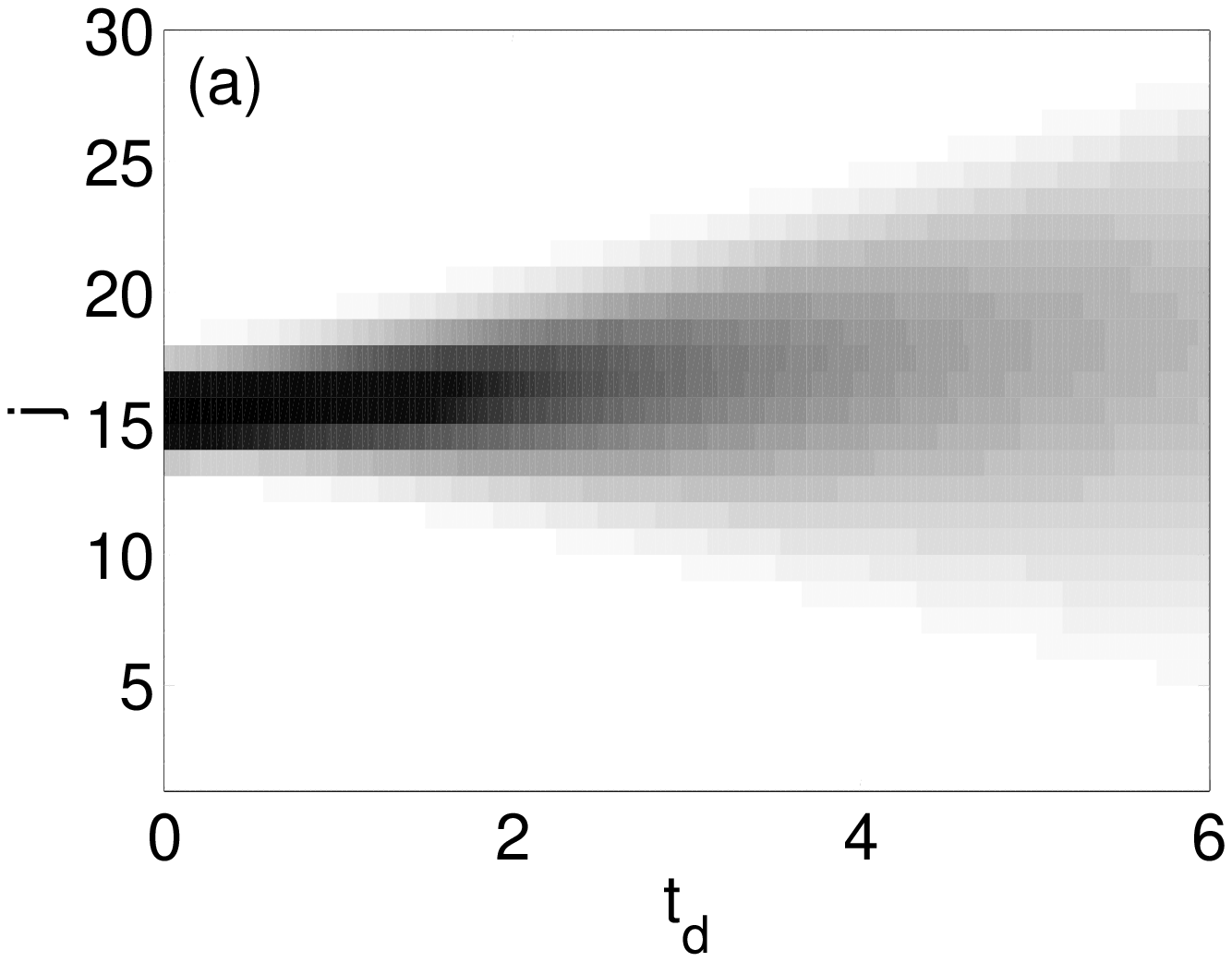}
  \includegraphics[width=4.3cm]{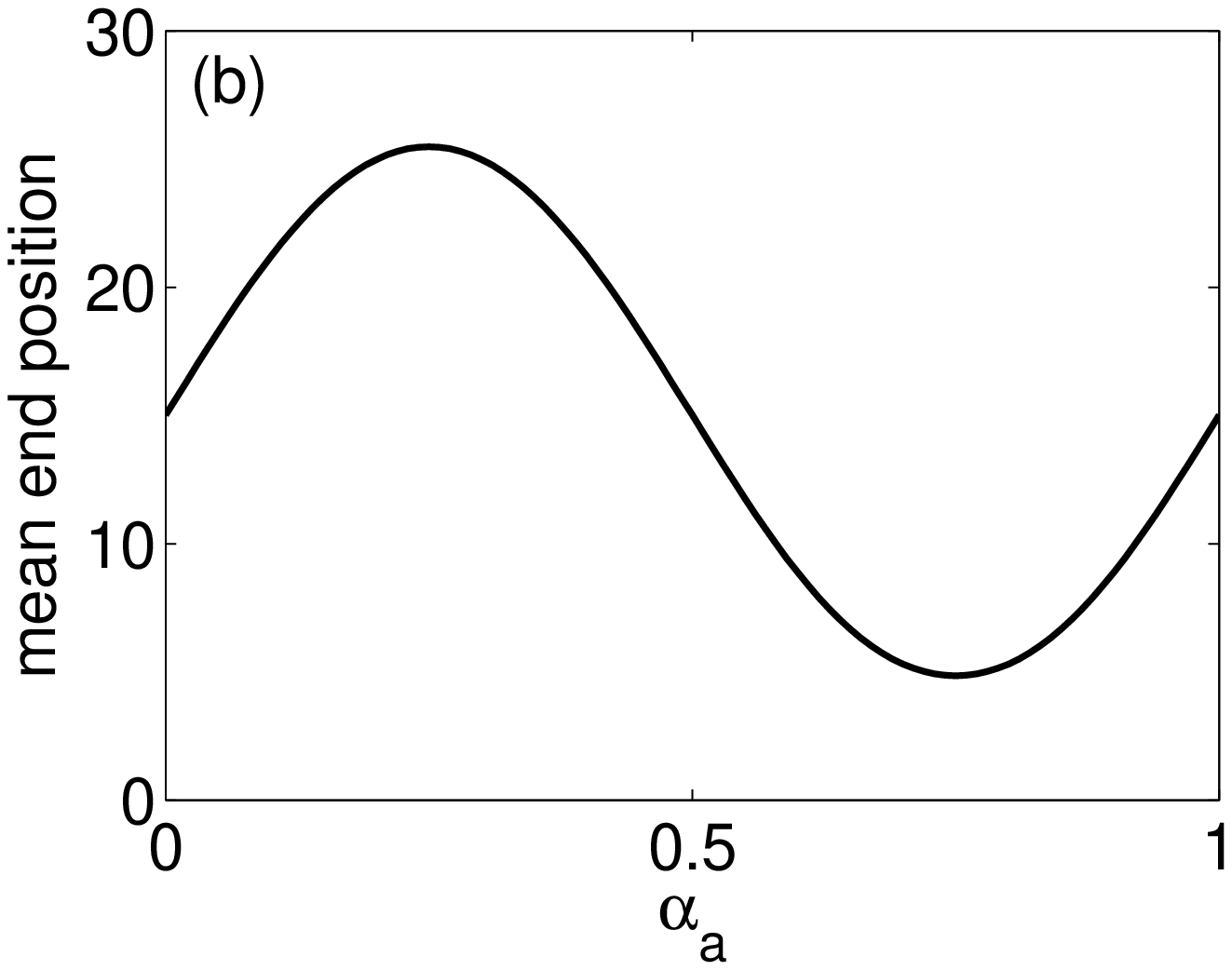}
\caption{\label{Fig:drift} (a) Plot of the impurity distribution
versus dimensionless time $t_d = t J_a/\hbar$. A darker colour
corresponds to a higher density. (b) Mean position of the impurity
after an evolution time of $t_d = 6$. We have chosen an initial
state $\ket{\psi}_\mathrm{ini}$ of width $w_\mathrm{ini} = \sqrt{2}$
centered at lattice site $j_0 = 15$. In (a), the chosen phase twist
was $\alpha_a = 0.03$.}
\end{figure}

This behaviour changes if a finite phase twist $\alpha_a$ is taken
into account as shown in fig.~\ref{Fig:drift}(a). For the chosen
parameters and a relatively short time of $t = 5 \hbar / J_a$, which
under standard experimental conditions corresponds to a few
miliseconds, a drift towards higher lattice site numbers is clearly
visible. Such drifts can be measured in experiments
\cite{Gupta-PRL-2005} and thus allow to probe the presence of an
induced phase twist. The mean position of the impurity after an
evolution time of $t = 5 \hbar /J_a$ is shown in
fig.~\ref{Fig:drift}(b). It exhibits a periodic behaviour in
$\alpha_a$ with a period of 1, which is caused simply by the fact
that the integer part of $\alpha_a$ only contributes a trivial
phase. In total, the behaviour is akin to an electron in a ring
subject to a magnetic flux. The electron will also drift towards a
preferential direction and thus will cause a current as known from
superconducting rings.

\subsection{Numerical simulations of the full ring system}

We now compare our prediction to a system where a numerically exact
solution is possible, namely a small two-mode Bose-Hubbard model. We
assume that the condensate is represented by a mode $c$ that
experiences a rotation whereas the other mode $a$ representing a
single impurity does not. The system is described by the Hamiltonian
\begin{align}
  \hat H_{BH} =& -J_a \sum_{\left\langle i,j \right\rangle}
   \hat a_i^\dagger \hat a_j + U_I
  \sum_j \hat a^\dagger_j \hat a_j \hat c^\dagger_j \hat c_j \\
  & -J_c \sum_{j} \left(\eh^{2 \pi \ii \alpha_c} \hat c^\dagger_j \hat
  c_{j+1} + \mathrm{h.c.} \right)
  + \frac{U_c}{2} \sum_j \hat c^\dagger_j \hat c^\dagger_j \hat c_j
  \hat c_j \,. \nonumber
\end{align}
Here, $J_c$ is the hopping matrix element of the atoms in mode $c$,
$U_c$ their interaction constant, $U_I$ the interaction between
atoms in mode $c$ and $a$, and $\alpha_c = m_c \Omega_c N_s a^2/2
\pi \hbar$, where $m_c$ is the mass of the atoms, $\Omega_c$ the
speed with which the ring for atoms in mode $c$ is rotated, $N_s$
the number of lattice sites, $a$ the lattice constant, and periodic
boundary conditions have been assumed \cite{Rey-PRA-2007}.

\begin{figure}
\centering
  \includegraphics[width=4.3cm]{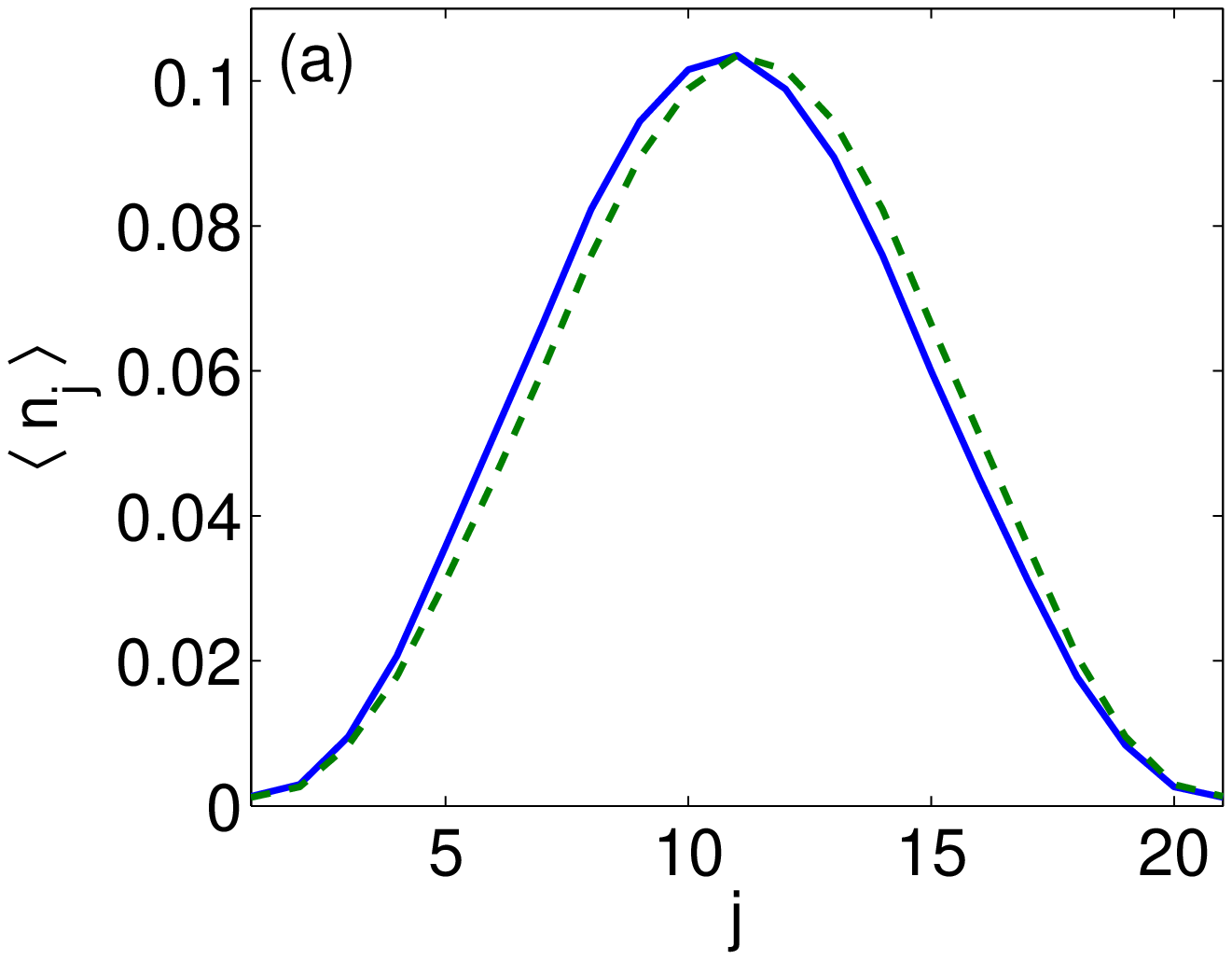}
  \includegraphics[width=4.3cm]{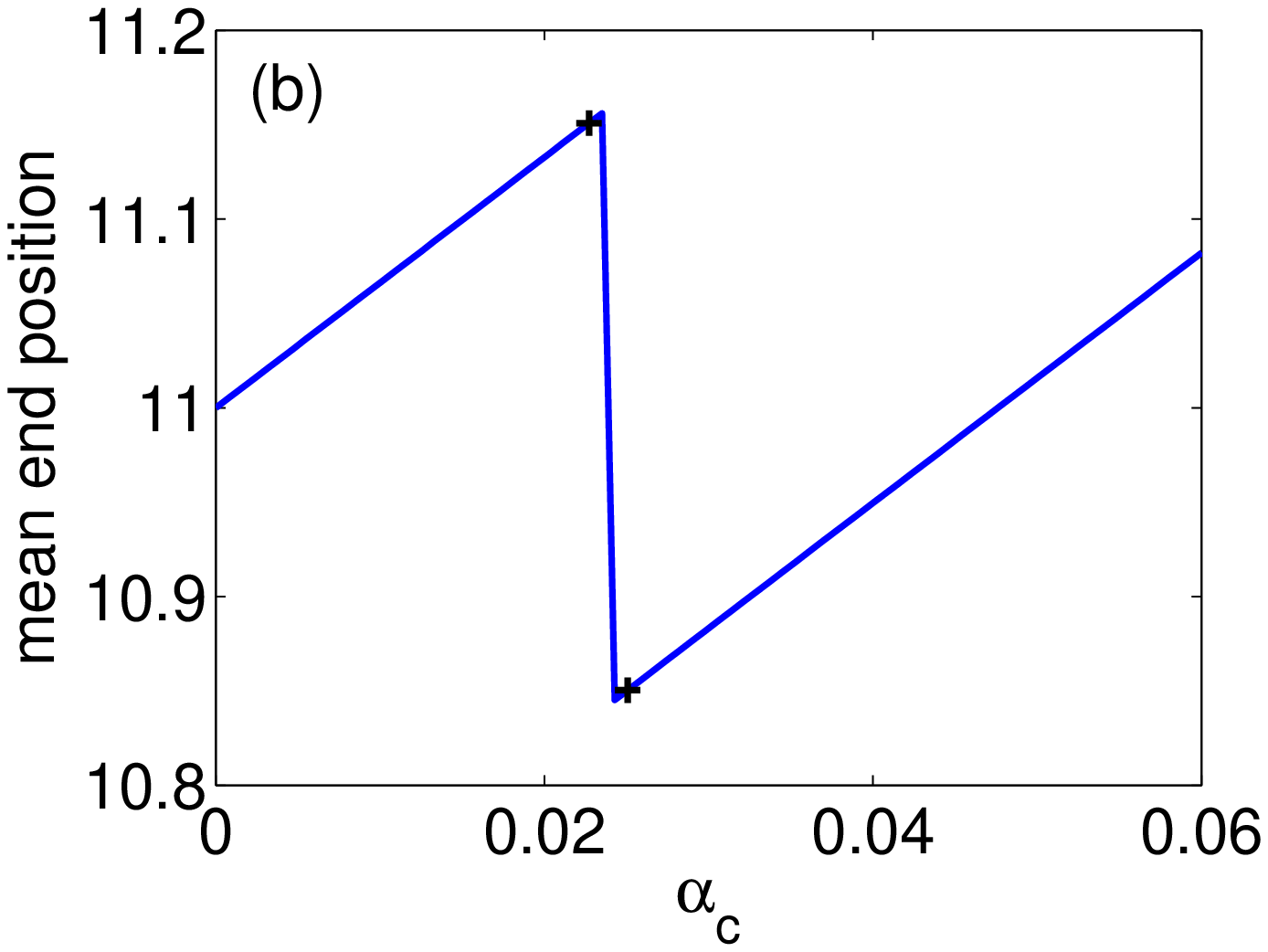}
\caption{\label{Fig:drift_num} Expansion of a single impurity atom
initially in state $\ket{\psi}_\mathrm{ini}$ with $j_0=11$ and width
$w_\mathrm{ini} = \sqrt{2}$. (a) Density distribution after a time
$t_d = t J_a/\hbar= 4$ for $\alpha_c = 0.0227$ (dashed line) and
$\alpha_c = 0.0251$ (solid line). (b) Mean position of the impurity
after the evolution time $t_d = 4$. The two crosses mark the
positions of the densities in (a). The other parameters are $U_c =
J_c = J_a$, $U_I = 2 J_c$ with three atoms in mode $c$ and a lattice
consisting of $N_s = 21$ sites.}
\end{figure}

We first calculate the ground state of the atoms in mode $c$ for
different values of $\alpha_c$ and vanishing coupling to the
impurity atom. We assume that the impurity is in state
$\ket{\psi}_\mathrm{ini}$ at time $t=0$ when the interaction $U_I$
is turned on and let the system evolve for a finite evolution time.
For two different $\Omega_c \propto \alpha_c$ the density
distributions of the impurity after this time are shown in
fig.~\ref{Fig:drift_num}(a), using typical experimental parameters.
A small drift to either the left or the right is clearly visible. We
expect this difference to be by far more pronounced for a larger
lattice with a longer evolution time. However, for computational
reasons we had to restrict ourselves to a small system size. The
behaviour of the mean end position versus the phase twist $\alpha_c$
is shown in fig.~\ref{Fig:drift_num}(b). As we see this drift
exhibits a jump at a value of $\alpha_c =1/2N_s$. The reason for
this jump is similar to the saw-tooth behaviour of the induced phase
$\alpha_a$ observed in fig.~\ref{Fig:alpha1Dana}(b). Due to the
quantisation of the BEC quasi-momenta there exist critical rotation
velocities $\Omega_c^\mathrm{crit} = 2 \pi^2 (2 j +1) \hbar/m_b
N_s^2 a^2$ at which the ground state of the BEC changes from one
quasi-momentum state to another one \cite{Rey-PRA-2007}. These
critical velocities correspond to phase twists $\alpha_c =
(2j+1)/2N_s$ for integer $j$. One can show that due to these jumps
the induced phase $\alpha_a$ changes analogously to the behaviour
shown in fig.~\ref{Fig:alpha1Dana}(b). This change is then also
reflected in the drifts that the impurity atoms experience.

\section{The two-dimensional condensate}

A phase twist $\alpha_{i,j}$ is also induced if we consider a
two-dimensional BEC in a trap rotating with angular velocity
$\Omega$. For simplicity we assume an interaction free condensate,
\textit{i.e.},~$g=0$. In the rotating frame, the (single particle)
Hamiltonian describing the condensate subject to a harmonic trapping
potential $\omega_\mathrm{tr}$ thus reads $ \hat H_0 = - (\hbar^2 /2
m_b) (\nabla')^2 +  m_b
 \omega_{\mathrm{tr}}^2 (r')^2/2 - \mu - \Omega \, \mathbf{e}_z \cdot  \mathbf{r}' \times
 \hat{\mathbf{p}}'$, where $\mathbf{e}_z$ is the unit vector
in $z$ direction and $\hat{\mathbf{p}}' = -\ii \hbar \nabla'$. It
can be shown that the solutions for this Hamiltonian are given by
$\psi_{n,l} (r, \varphi) = \sqrt{n!/\pi (n+l)!} \,
  \eh^{\ii l \varphi} \eh^{-\xi^2/2} \xi^{l} L_n^l(\xi^2) /a_0$ with
  energies
  $E_{n,l} + \mu = (2n + l + 1)\hbar \omega_{\mathrm{tr}} -
\hbar \Omega l$. Here, $\xi = r/a_0$ with $a_0 = \sqrt{\hbar / m_b
\omega_\mathrm{tr}}$ the harmonic oscillator length and
$(r,\varphi)$ are polar coordinates. The quantum numbers $n$ and $l$
are restricted to $n = 0,1,2,...$ and $l = -n, -n+1, ...\,$. From
this we see that for high rotation speeds $|\Omega| >
\omega_\mathrm{tr}$ the system is unstable since for suitable
$(n,l)$ the energy can be made infinitely small which corresponds to
the atoms being driven infinitely far away from the origin due to
the centrifugal term. We thus restrict the rotation speed to
moderate values $|\Omega| < \omega_\mathrm{tr}$, for which the
ground state is always given by $(n,l) = (0,0)$.

For brevity, let us introduce the multi-index $\nu = (n,l)$. If we
assume that the condensate is in a state $\nu_0$, we can expand the
field operator of the condensate as $\hat \psi(r,\varphi) \to
\sqrt{N_0} \psi_{\nu_0} (r, \varphi)
  + \sum_{\nu \neq \nu_0} \psi_\nu(r,\varphi) \hat b_\nu $, where
$\langle \hat b_{\nu_0}^\dagger \hat b_{\nu_0}\rangle = N_0$ is the
number of atoms in state $\nu_0$, and we assume that $\langle \hat
b_{\nu}^\dagger \hat b_{\nu}\rangle \ll N_0$ for $\nu \neq \nu_0$.
As we consider the interaction-free case this is exactly the same as
an expansion in terms of Bogolibov phonons, which would yield $v_\nu
= 0$ and $u_\nu = \psi_\nu(r,\varphi)$. The chemical potential is
determined by solving the Gross-Pitaevskii equation for the
corresponding state $\nu_0$, which gives $\mu_{\nu_0} =
-E_{n_0,l_0}$. This allows us to derive the phase twist
$\alpha_{i,j}$ in the same way as for the ring optical lattice. In
order to solve the integral in the expression for the coupling
matrix elements $M_{j,\nu}$ we assume that the Wannier functions are
strongly localised and can be approximated by a Dirac delta
function. If the condensate is in the ground state $\nu_0 = (0,0)$,
we find for the induced phase twist
\begin{equation}
\begin{split}
  \alpha_{i,j}^{(0,0)}  =\,& \frac{\kappa^2 N_0}{2 \pi^3 a_0^4} \sum_{\nu \neq
  \nu_0} \frac{1}{(\hbar \omega_\nu)^2} \frac{n!}{(n+l)!} \eh^{-\xi_i^2 -
  \xi_j^2} \\
  &\times \xi_i^l \xi_j^l L_n^l(\xi_i^2) L_n^l(\xi_j^2) \sin(l(\varphi_i -
  \varphi_j)) \,,
\end{split}
\end{equation}
where $(\xi_i, \varphi_i)$ and $(\xi_j, \varphi_j)$ are the
coordinates of the two lattice sites before and after the jump of
the impurity. We immediately see that if $\varphi_i = \varphi_j$,
which corresponds to a hopping in radial direction, the induced
phase is zero. Furthermore, the induced phase vanishes as $\xi_i$ or
$\xi_j$ go to infinity, which can be explained by the vanishing
density of the condensate for large distances.

\begin{figure}
\centering
  \includegraphics[width=4.3cm]{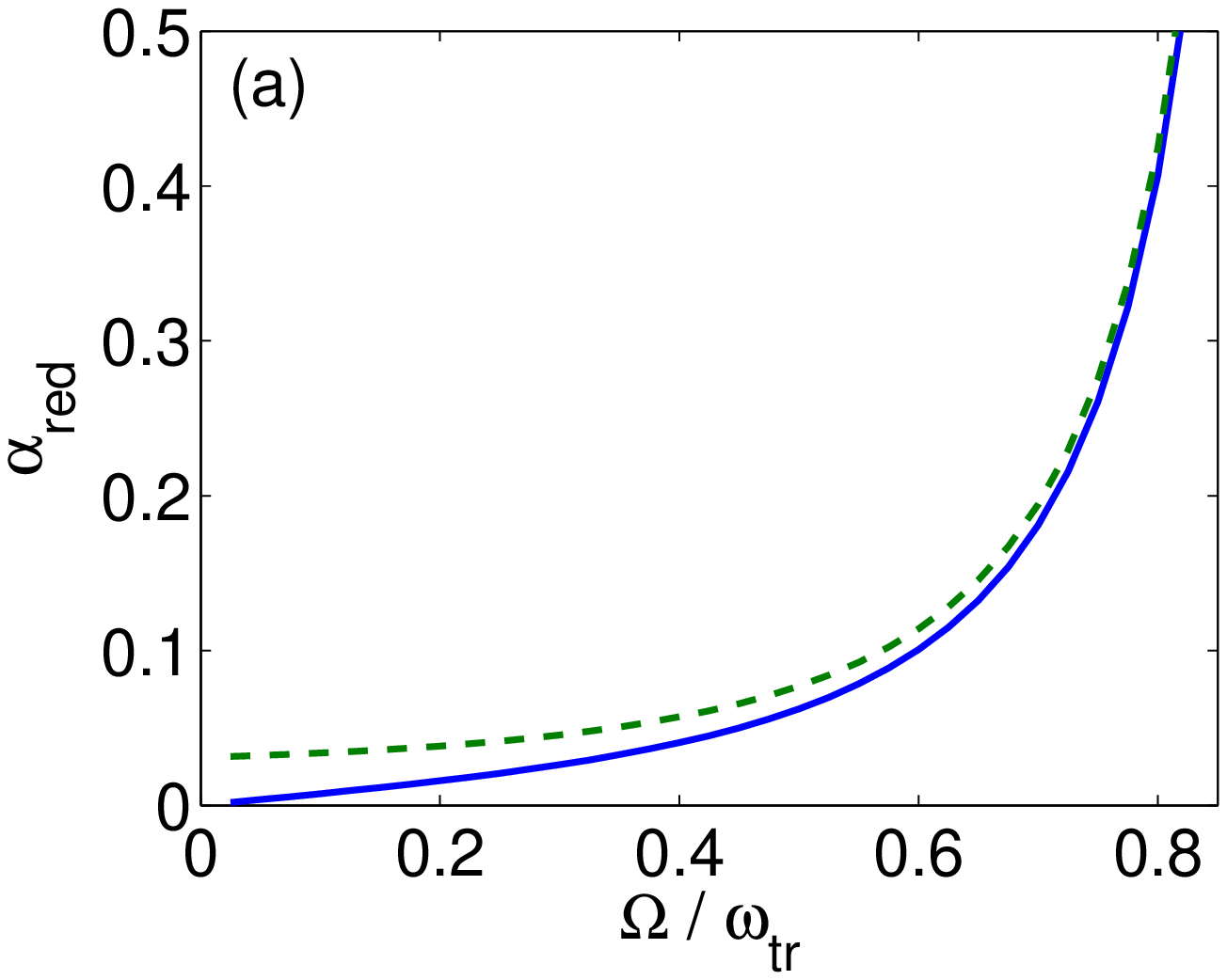}
  \includegraphics[width=4.3cm]{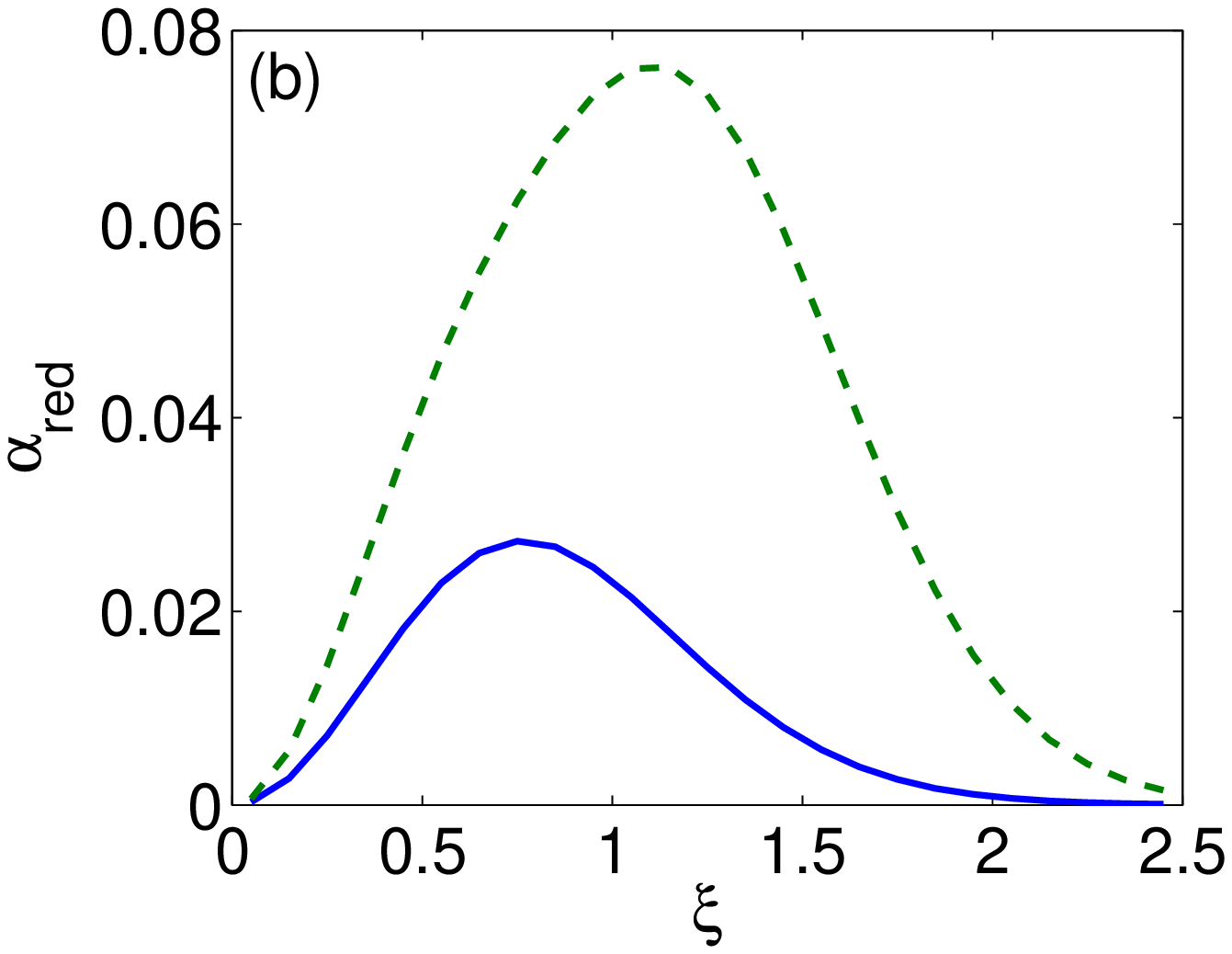}
\caption{\label{Fig:2Dphase} (a) Reduced phase twists
$\alpha_\mathrm{red}^{(n,l)} = \alpha_{i,j}^{(n,l)} \times (\kappa^2  
N_0/2 \pi^3 a_0^4 (\hbar \omega_\mathrm{tr})^2)^{-1}$ for the ground
state $(n,l)=(0,0)$ (solid line) and the vortex state $(n,l)=(0,1)$
(dashed line) versus dimensionless angular rotation speed
$\Omega/\omega_\mathrm{tr}$. Here, $\xi_i = \xi_j = 0.5$ and
$\varphi_i - \varphi_j = 0.1$. (b) Reduced phase twists versus $\xi
= \xi_i=\xi_j$. The angular velocity is given by $\Omega = 0.25
\omega_\mathrm{tr}$ and $\varphi_i - \varphi_j = 0.1$.}
\end{figure}
%
%

Our theory is also applicable if the condensate is initially in an
excited state, for example in the state $\nu_0 = (0,1)$ that
exhibits a vortex. In this case, we derive for the phase twist
\begin{equation}
\begin{split}
  \alpha_{i,j}^{(0,1)} =\,& \frac{\kappa^2 N_0}{2 \pi^3 a_0^4} \sum_{\nu \neq
  \nu_0} \frac{1}{(\hbar \omega_\nu)^2} \frac{n!}{(n+l)!}
  \eh^{-\xi_i^2 -\xi_j^2}   (\xi_i
  \xi_j)^{l+1} \\
  &\times L_n^l(\xi_i^2) L_n^l(\xi_j^2)
  \sin[(\varphi_i - \varphi_j)(l+1)] \,.
\end{split}
\end{equation}

The behaviour of the two phase twists for varying angular rotation
speed $\Omega$ is shown in fig.~\ref{Fig:2Dphase}(a). In both cases,
the induced phase twists increase as the rotation speed $\Omega$ is
increased and diverge as $\Omega$ approaches the trapping frequency
$\omega_\mathrm{tr}$. We furthermore observe that for small rotation
frequencies the state exhibiting a vortex still induces a finite
phase twist, whereas $\alpha_{i,j}^{(0,0)}$ vanishes as $\Omega$
goes to zero. The dependence of the phase twist on the distance from
the origin is shown in fig.~\ref{Fig:2Dphase}(b). For large
distances $\xi$, the phase twist is zero due to the vanishing
condensate density. For small distances the phase twist vanishes as
well, due to the fixed difference $\varphi_i - \varphi_j$. The
spatial dependence of $\alpha_{i,j}^{\nu_0}$ leads to the
realisation of an artificial magnetic field, which in general will
be non-homogenous. We expect that the spatial dependence of this
field can be controlled by varying the trapping potential and thus
the density of the BEC. The behaviour of the impurities can also be
used to probe the state of the condensate. These effects will be the
topic of future research.

\section{Summary}

We have investigated an alternative way to create artificial
magnetic fields in a system of ultracold atoms in an optical
lattice. We have shown analytically that for impurities trapped in a
ring-shaped optical lattice submerged into a ring-shaped, rotating
Bose-Einstein condensate a drift in the transport of the impurities
occurs. This drift was qualitatively confirmed by a full numerical
simulation of a two-mode Bose-Hubbard model. We also showed that an
artificial magnetic field can be created in a two-dimensional
optical lattice by submerging it into a moderately rotating,
harmonically trapped condensate. With this, our system lends itself
as a well-suited quantum simulator for investigating phenomena
encountered in condensed matter physics.

\acknowledgements

The authors thank M.~Bruderer, S.~R.~Clark, and M.~Rosenkranz for
helpful discussions. This work was supported by the United Kingdom
EPSRC through QIP IRC (Grant No.~GR/S82176/01) and EuroQUAM Project
No.~EP/E041612/1, the EU through the STREP project OLAQUI, and the
Keble Association (A.K.).

\bibliographystyle{eplbib}
\bibliography{../../Biblio}

\end{document}